\documentstyle[12pt]{article}
\newcommand{\br}{\mbox{\boldmath $r$}}

\newcommand{\bepsilon}{\mbox{\boldmath $\epsilon$}}

\newcommand{\ba}{\begin{eqnarray}}
\newcommand{\ea}{\end{eqnarray}}
\newcommand{\be}{\begin{equation}}
\newcommand{\ee}{\end{equation}}
\newcommand{\trom}{{\rm \,  tr \,}} 
\begin{document}
\vspace*{1.truein}
\centerline{\bf MESON PHOTOPRODUCTION AT THRESHOLD } 
\centerline{\bf IN THREE FLAVOR SOLITON MODELS }
\vspace*{0.035truein}
 
\vspace{0.37truein}
\centerline{\footnotesize B. SCHWESINGER 
} 
\vspace*{0.015truein}
\centerline{\footnotesize\it FB Physik, University Siegen, Postfach 101240} 
\baselineskip=12pt
\centerline{\footnotesize\it  D-57068 Siegen} 
\vglue 12pt
\baselineskip=12pt
\baselineskip=6mm
\vglue 12pt
\vspace{1.truein}
\abstract{In Skyrme-type models, the leading term of the low-energy 
photoproduction amplitude is identical to the standard expresssion and 
independent of the number of flavors considered, but subleading terms
are not. }

\vspace{3.truein}

{\footnotesize SI-96-TP3S1} \\
\vspace{.015truein}
{\footnotesize hep-ph/9608268}
\eject

Recently it has been shown, that a wide class of topological
soliton models within $SU(2)$ lead to the well-known\cite{Baenst} 
low-energy theorem of pion
photoproduction derivable up to ${\cal O}(N_C^{-1})$ and ${\cal
O}(m_\pi)$, ref.\cite{skylet}. Generalization of such models to more
flavors is quite straight-forward in Skyrme-type models and much effort
has been spent, for instance, on the properties of baryons rotating in
$SU(3)$, see e.g. ref.\cite{RRA}. In
this little note I wonder, whether the low-energy theorem is altered, when
further flavors are added to the $SU(2)$ soliton. Also, it might be
possible to make additional
statements on meson production channels beyond the four traditional
pion-nucleon  channels, when the photon either strikes a non-strange
baryon or produces strangeness in the meson-baryon exit channel.\\

In soliton models the meson-baryon-photon interactions follow
exclusively from the vector and axial vector currents inherent in the
purely mesonic Lagrange density ${\cal L}(U,\partial_\mu U)$. 
These currents are defined as
\be
V_a^\mu (U) = -i \trom \frac{\delta {\cal L}}{\delta \partial_\mu U}
[ \frac{\lambda_a}{2} , U ] \, , \qquad
A_a^\mu (U) = -i \trom \frac{\delta {\cal L}}{\delta \partial_\mu U}
\{ \frac{\lambda_a}{2} , U \} \, .
\ee
For definiteness I also list the currents from the chiral anomaly 
\ba
V_a^\mu (U) = \frac{N_C}{48 \pi^2} \epsilon^{\mu \nu \rho \sigma}
 \trom 
{\lambda_a 
\over{2}}\left\{- U^\dagger \partial_\nu U
U^\dagger \partial_\rho U U^\dagger \partial_\sigma U
+ U \partial_\nu U^\dagger  U \partial_\rho U^\dagger 
 U \partial_\sigma U^\dagger \right\}\\
A_a^\mu (U) = \frac{N_C}{48 \pi^2} \epsilon^{\mu \nu \rho \sigma}
 \trom {\lambda_a\over 2} \left\{ ~~U^\dagger \partial_\nu U
U^\dagger \partial_\rho U U^\dagger \partial_\sigma U
+ U \partial_\nu U^\dagger  U \partial_\rho U^\dagger 
 U \partial_\sigma U^\dagger \right\}\nonumber
\ea
If the parameters $\phi$ of a chiral rotation
\be
U=e^{\frac{i}{2} \phi \cdot \lambda} \, U_0 \, e^{\frac{i}{2} \phi \cdot
\lambda} 
\ee
are introduced, the following identity can be shown to hold for
arbitrary $U_0$
\be
 V_a^\mu (U) = V_a^\mu (U_0) - f_{abc}\, \phi_b \, A_c^\mu (U_0)
+ {\cal O}(\partial \phi, \phi^2)
\ee
for all parts of the current  originating from chirally symmetric
terms of the action.  

In Skyrme-type models, the nucleon is based on 
topological field configurations of the hedgehog soliton embedded in
the isospin subgroup $SU(2) \subset SU(3)$ 
\be
U_H(\br)= e^{i \tau \cdot \hat r \chi(r) } \, .
\ee
Kinematical and spin degrees of freedom are
introduced by collective coordinates. For small velocities such
coordinates can originate from a
Galilean transformation of the center of mass ${\bf r} \rightarrow
{\bf r+ X}$
or from adiabatically slow flavor rotations $A\, U_H \, A^\dagger$.  

In the
low-energy region of interest here, meson-baryon S-wave configurations 
can elegantly be written as 
\be
U =  e^{\frac{i}{2} \lambda \cdot \phi} \,A \, \, 
U_H \Bigl( x_i + \frac{3 g_A}{2 M} \phi_a D_{ai}(A) \Bigr) \, \,A^\dagger \,
e^{\frac{i}{2} \lambda \cdot \phi } \, .
\ee 
where at threshold a meson is introduced as the term linear in the
angles $\phi$ of a chiral rotation.
Actually, it represents a linear combination of a 
chiral rotation of the hedgehog and a translation,
$X_i=\frac{3 g_A}{2 M} \phi_a D_{ai}(A)$. (As usual we use 
$D_{ab}(A)= \frac{1}{2} \trom \lambda_a A^\dagger \lambda_b A$. The
index $i\in\{1,2,3\}$ stands for the three spatial directions).
In case of chiral symmetry, i.e. when chiral symmetry breaking
mass terms
are absent, both modes, translation and chiral rotation 
are zero frequency solutions to the adiabatic
equations of motion for small amplitude fluctuations. The special
linear combination given here is determined by the fact, that
the S-wave scattering solution is orthogonal on the 
localized, purely translational zero mode. The overlap integrals
of infinitesimal fluctuations between chiral rotations and translations
involve the mass $M=-L[U_H]$ 
of the soliton and its axial coupling $g_A$,
\ba
\int A^j_a\left( A U_H A^\dagger \right) d^3r 
&=& D_{ab}(A) \int A_b^j \left( U_H \right) d^3r \\
&=&
\left\{ \begin{array}{ll}
-\frac{3}{2} D_{aj}(A)  \, g_A^{SU(2)} &\, {\rm for} A\in SU(2) \\
-{30\over{14}} D_{aj}(A)  \, g_A^{SU(3)}&\, {\rm for} A\in SU(3)
\end{array} \right. \quad .\nonumber
\ea 
The proton-neutron matrix elements of the axial current define the
isovector axial charges and from this definition we deduce
for $SU(3)$-symmetric nucleon states
\be
g_A^{SU(3)} = {7\over{10}} g_A^{SU(2)}, \qquad g_A^{SU(2)} =  g_A \, .
\ee
The low-energy production amplitudes now follow directly from the
lagrangian linear in the photon $a_\mu$ and the meson field $\phi$. 

\ba
L_{\gamma \phi} = L_{\gamma \phi}^V + L_{\gamma \phi}^{ano} =
-\mid e \mid \int a_\mu \left [ V^\mu_e(U) \right ]_{lin} d^3r
\ea
As usual, we abbreviate $(e)= (3) + \frac{1}{\sqrt{3}} (8)$ for 
electromagnetic charge indices. 
At low energies, in Coulomb gauge, only
the constant polarization vector ${\bf a} = -\frac{1}{4 \pi} \bepsilon$
is important.
In contrast to the $SU(2)$-case,
the contribution of the anomaly in $SU(3)$ leads to an additional
vector current associated with the eighth generator, whereas the
$SU(2)$ expression involves the topological winding number current.
The main production amplitude at threshold, as in $SU(2)$, comes from
the isovector charge of the mesons, however, and may easily be obtained
via eqs.(4,7) as
\ba
L_{\gamma \phi}^V  = -\frac{\mid e \mid }{8 \pi } 
 3 g_A \, f_{ebc} \, \phi_b \, D_{ci}(A) \,\epsilon_i \,.
\ea
The only difference relative to the two flavor case is the appearence
of $f_{ebc}$ instead of $\epsilon_{3bc}$ and the $SU(3)$-D-function,
the nucleon matrix elements of which are changed from 
$-{1\over 3} \tau_c \sigma_i\stackrel{SU(3)}{\rightarrow}
                                     -{7\over 30} \tau_c \sigma_i$
for $SU(3)$ symmetry; since the differing factor of $7\over10$ is
absorbed into the definition
of the nucleon axial charge in the three flavor case, eq.(8), this
isovector amplitude is identical to its two-flavor partner when pions
are produced on non-strange baryons. Checking out other mesons and
baryons we find that only charged mesons can be produced here, $\pi_0,
\eta, K_0, \bar K_0$ do not couple in this lowest order
Kroll-Ruderman amplitude.

Corrections linear in the meson mass arise 
for winding number $B=1$ from the anomalous current. In contrast to the
two flavor case there are two different contributions, the first one
being an immediate generalization of the two-flavor isoscalar current 
\ba
L_{\gamma \phi} ^{(ano,1)}  =  \frac{\mid e \mid }{8 \pi}
\frac{ 3 g_A}{2 M} \, \dot{\phi_b} D_{bi}(A) \epsilon_i \, B 
\,\frac{N_C}{\sqrt{3}} D_{e8}(A)
\ea
as may be checked by taking the $SU(2)$-limit $D_{38}(A)\rightarrow 0,
D_{88}(A)\rightarrow 1$. In $SU(3)$ this term
allows for the production of neutral mesons just as the other, novel term
does (we abbreviate $k,k' \in \{4,5,6,7\}$):
\ba
L_{\gamma \phi}^{(ano,2)} =  \frac{\mid e \mid }{8 \pi}
\frac{N_C}{9} d_{ikk'} D_{ek}(A) \dot{\phi_b} D_{bk'}(A) \, \epsilon_i
\, \int r\,s\, B^0(U_H) d^3r. 
\ea
The novel term is no longer
expressible by the charges $g_A, B $ of the baryon but involves an
integral over the winding number density $B^0$ of the soliton together
with  a trigonometric function of the chiral angle ($r\,s=r\,\sin{\chi(r)}$).

The three different production amplitudes, eqs.(10,11,12), constitute
all possible terms in $SU(3)$-symmetric soliton models up to linear
order in the meson masses and zeroth order in the rotational velocities
of the soliton. How useful is such a classification? In soliton models 
$SU(3)$ symmetry
is broken by two different kinds of terms:\\
(i) meson mass terms which split
the $K$ and $\eta$ masses from the pion. Since they involve no
gradients of the chiral fields, they do not lead to additional
photocouplings, their sole effect being a distortion of the hedgehog
fields and the chiral rotation around them. Their contributions are always
quadratic in the meson masses with exception of the time dependence of
the produced (fluctuating) mesons, which at threshold
also has a linear term because of the
quantization of the fluctuations. 
We will examine the
implicit effects of this flavor and chiral symmetry breaking on the threshold
amplitudes.\\ 
(ii) Kinetic symmetry breakers,
which lead to different weak decays of the mesons, is the other kind of
terms. They will lead to further photocouplings. Generally, they are
numerically less important and will be omitted in the present
considerations.

It is clear that the $K$ and $\eta$ masses are not small numerically,
thus the amplitudes in eqs.(10,11,12) are not trustable numerically, but
they represent a systematic expansion of the threshold amplitudes,
which certainly is interesting in itself, and it covers the whole octet
of baryons in the incoming channel together with the octet of baryons
and mesons in the exit channels leading to a multitude of different
reactions relative to only four $SU(2)$-cases. The latter are, however,
more trustable, because pion masses are numerically small. We
therefore will start with the pion-nucleon amplitudes and their
modifications when going to  SU(3).

\eject

{\bf Pion production in $SU(3)$}\\[.3cm]

\begin{table} [h]
\begin{center}
\parbox{12cm}{
\caption[]{\footnotesize Kroll-Ruderman amplitudes  in units 
$10^{-3} m_{\pi^+}^{-1}$ for the cases: \\
(i) the standard low-energy theorem in $SU(2)$.\\
(ii) $SU(2)$ soliton model, ref.\cite{skylet}.\\
(iii) $SU(3)$ soliton model according to eqs.(10, 11, 12).\\ 
(iv) reanalysed experimental data, M: ref.\cite{Mainz}, S: ref.\cite{Saclay}.} 
}

\begin{tabular}  {| l| r r r | r |}
\hline
           & standard    &  SU(2) soliton  &  SU(3) soliton& experiment    \\  
           &   LET~~~    &    model~~~~    &   model~~~~ &        \\
\hline
& & & &   \\  
$A^{n(\gamma , \pi^- ) p }$ & $-$31.8 & $-$31.8  & $-$31.3 
                                              & $-$31.4$\pm$1.3$^M$ \\
                            &         &   &         & $-$32.2
                                                 $\pm$1.2$^S$ \\
                            &         &          &         &                 \\
$A^{p(\gamma , \pi^+ ) n }$ & $-$27.4 & $-$27.4  & $-$27.9 & 
                                                         $-$27.9$\pm$0.5$^M$ \\
                            &         &   &         & 
                                                        $-$28.8$\pm$0.7$^S$ \\
                            &         &          &         &                 \\
$A^{p(\gamma , \pi^0 ) p }$ &  $-$2.5 &  $-$1.6  & $-$.8   &  
                                                          $-$2.0$\pm$0.2$^M$ \\
                            &         &    &         &
                                                          $-$1.5$\pm$0.3$^S$ \\
                            &         &          &         &          \\ 
$A^{n(\gamma , \pi^0 ) n }$ &   0.4   &  1.6     & .6   &       \\ 
                            &         &      &         &         \\
           &           &           &            &             \\
\hline
\end{tabular}
\end{center}

\end{table}
Table 1. shows the numerical effects of symmetrically adding the strange flavor
degree of freedom to the rotating soliton. In this case the strangeness
content of non-strange baryons is, of course, maximal ($\frac{7}{30}$).
Because the
leading term for charged pion amplitudes is independent from the number of
flavors involved, deviations only arise from the subleading terms and
are thus most easily noticeable in the amplitudes for neutral pion
production. Of course, the assumption of $SU(3)$-symmetry is not given
in the baryonic wave functions, see e.g. ref.\cite{RRA}, 
and here it only serves as a
demonstration  for the fact, that $SU(3)$ leads to a different and
noticeable change coming from the anomalous terms of the action.
Incidentally, the novel term $L_{\gamma \phi}^{(ano,2)}$, eq.(12), only
amounts to roughly $10\%$ of $L_{\gamma \phi}^{(ano,1)}$, eq.(11) and
is unimportant in this respect. The latter, on the other hand, continuously
returns to its $SU(2)$ value with decreasing strangeness content of the
nucleons. I have already sketched this limit in the remark 
following eq.(11).\\[.3cm]

\eject
{\bf Strangeness production}\\[.3cm]

Turning to the photoproduction of strangeness on the nucleon we are
leaving safe ground, because several reasons lead to numerically uncontrolable
statements. Nevertheless, the amplitudes derived allow for some
qualitative insight.

\begin {table} [h]

\begin{center}
\parbox{9.5cm}{
\caption[]{\footnotesize The Born coupling  constants for reactions with strangeness
production as given by three flavor soliton models from eq.(10).
The first column shows the case for $SU(3)$-symmetry of the baryon wave
functions, the second column uses realistic symmetry breaking for the
baryons, and the last column shows the effect of a formfactor as
explained in the text.  
Quantities with a '*' are normalized to their empirical value.
} }

\begin{tabular}  {| l| r r r | }
\hline
           & SU(3)~~~~~  &  broken   &  FF~~~~      \\  
           & symmetry   &   SU(3)~~&   effects         \\
\hline
& & &    \\  
$g_{\pi NN}/\sqrt{4 \pi}  $ & 3.78$^*$       & 3.78$^*$    & --    \\
                            &         &          &          \\ 
$g_{K \Sigma N}/\sqrt{4 \pi}  $ & $-$1.08 & $-$.89  & $\sim$0     \\
                            &         &          &          \\
$g_{K \Lambda N}/\sqrt{4 \pi}  $ & 3.74   & 2.69    & $\sim$0     \\
                            &         &          &          \\
\hline
\end{tabular}
\end{center}

\end{table}

Table 2. displays the Born couplings, as they can be deduced from
soliton models taking Euler-angle matrixelements over eq.(10) for different
baryons. For $SU(3)$-symmetry, first column, they coincide with conventional
expresssions, ref.\cite{borncoupl}.
These couplings have been considered as being far too large, to
accomodate for the strange\-ness production data \cite{MBA}, and other
strategies have then been attempted there.

The soliton model allows for more
sophisticated approaches:\\
(i) one may relax the assumption of
$SU(3)$-symmetry on the baryon wave functions and use those
diagonalizing the $SU(3)$-symmetry breaking. Using e.g. the rigid
rotator approximation\cite{RRA} one may see from the
second column of table 2., that this leads to a reduction of the born
couplings.\\
(ii) the soliton model offers, in a natural way, formfactors
to these couplings: the axial current  in eq.(7) only leads
to the axial charge in eq.(10) when integrated as in eq.(9) with a 
constant photon field. Inclusion of non-zero 
wavenumbers $k$ of the photon introduces an
additional $j_0(kr)$ into the integral, leading to $g_A(k)$.
Of course, this k-dependence is beyond the order of the amplitudes
restricted to here, i.e. ${\cal O}(m_K)$, but numerically most
important: with this inclusion the strange Born couplings are all
totally suppressed at kaon threshold, column 3. in table 2. 
Similar things will happen 
for the other two contibutions, eqs.(11,12), not considered here.

The conclusion from the soliton model is thus, that the expansion of
strangeness production amplitudes is not possible in the way attempted here and
much more effort must be invested. I will briefly sketch this effort: 
the assumption that the kaon fluctuation used here is a 
constant chiral rotation at $K \Lambda$ or $K \Sigma$ threshold is 
not justified. In meson-baryon scattering these channels plus
additional ones, predominantly $\pi N$, $\eta N$ all couple with each
other. In the, to my knowledge, only existing scattering
calculation\cite{mbscatt} in soliton models with broken $SU(3)$ I had
suggested that $K \Lambda$, $K \Sigma$ and $\eta N$ form a bound state
below $\eta N$ threshold. (Closing
the  $\pi N$-channel artificially I can actually calculate such a bound
state at roughly 450MeV above the nucleon mass). If true, such a complex
bound state way below the $K \Lambda$ or $K \Sigma$ threshold must
strongly distort the kaonic fluctuations to something far from the
constant chiral rotation I attempted to use here at threshold. Therefore, the
photoproduction amplitudes for strangeness production can only be
obtained after solving the meson-baryon scattering problem then inserting
the obtained meson waves into the one photon production vertex as has
been done for the case of $SU(2)$, refs.\cite{ppsu2}

Incidentally, this bound state  explains 
the known strong $\eta$ decay branching of the S11(1535)
resonance\cite{mbscatt}  (and similar ones with higher
strangeness, e.g. $K \Xi$ in $\bar K N$- scattering). The
$\pi$ baryon, $\eta$ baryon channels have practically no direct coupling in
soliton models. A similar statement has been made recently\cite{Weise}
in connection with chiral perturbation theory.\\[.3cm]

{\bf Eta production}\\[.3cm]

Almost identical reasons as put forward in the case of strangeness production
apply to $\eta$ production: a constant chiral $\eta$-rotation cannot be
sufficient in the presence of strongly coupled bound meson-baryon
channels 
($K \Lambda$, $K \Sigma$ and $\eta N$) 
already present below $\eta N$ threshold.\\[.3cm]

{\bf Conclusion}\\[.3cm]

In solitonic approaches the low-energy pion S-wave 
is given explicitly as
a constant chiral rotation orthogonal on the soliton translation.
The well-known low-energy theorem for pion photoproduction, following
from this fact  acquires a 
slightly different form in the subleading terms (production of neutral
pions) when strangeness degrees of freedom are included for mesons and
the soliton. 

Low-energy theorems for pion production on strange baryons are then also 
derivable and numerically trustable.

Production amplitudes for the heavier mesons obtained by the 
same method as for pions are not trustable, because the S-wave
resonances strongly
violate the assumption of the heavy meson wave being a chiral rotation
at threshold. I have argued, that this violation and the $\eta$-decay
of some S-wave resonances are closely connected due to the existence of
bound states of strangeness +1 kaons with strange baryons.\\[.3cm]

{\bf Acknowledgements}\\[.3cm]

Without the very personal, generous and selfless help of Gottfried
Holzwarth many things
would have proved impossible: this work certainly couldn't have
been completed. I feel and owe deep gratitude to him.

I thank Hans Walliser for critically accompanying this work,
recalculating part of the algebra, 
pointing out a missing term, and proofreading.



\end{document}